\pdfoutput=1

\documentclass[twocolumn,english,aps,superscriptaddress,prb]{revtex4-2}

\usepackage{babel}
\usepackage{amsmath}
\usepackage{amssymb}
\usepackage{graphicx}

\newcommand{\appropto}{\mathrel{\vcenter{
  \offinterlineskip\halign{\hfil$##$\cr
    \propto\cr\noalign{\kern2pt}\sim\cr\noalign{\kern-2pt}}}}}

\begin{document}

\title{Realization of Wess-Zumino-Witten transitions with levels $k=6$ and $k=4$ in a frustrated spin-3 chain}

\author{Natalia Chepiga}
\affiliation{Kavli Institute of Nanoscience, Delft University of Technology, Lorentzweg 1, 2628 CJ Delft, The Netherlands}

\date{\today}
\begin{abstract}
We study dimerization transitions in a frustrated spin-3 chain with next-nearest neighbor and three-site interactions. We show that two independent coupling constants of the model are sufficient to fine-tune the system to the critical point in the Wess-Zumino-Witten SU(2)$_6$ universality class. This critical point appears as the end point of an extended SU(2)$_4$ critical line. This implies that the renormalization group flow lead to the critical theory with the largest level $k$ such that the number of relevant operators is reduced by one and the parity of the level is preserved. Furthermore, we argue that due to the presence of marginal operator there is only one point in the SU(2)$_6$ universality class.
In addition, we report the appearance of non-magnetic Ising transition between the topologically trivial uniform and dimerized phases. This transition takes place within the singlet sector, while magnetic gap remains open.
\end{abstract}

\maketitle

\section{Introduction}
Antiferromagnetic Heisenberg spin chains have attracted a lot of attention over the years. Competing interactions induce frustration and are known to lead to new phases and quantum phase transitions. 
For example, the $J_1-J_2$ spin-1/2 chain undergoes a Kosterlitz-Thouless transition\cite{Kosterlitz} into a spontaneously dimerized phase\cite{MajumdarGhosh,okamoto}. By contrast, phases realized in the $J_1-J_2$ spin-1 chain are non-dimerized and gapped: for $J_2/J_1\lesssim 0.75$ the chain is in the topologically non-trivial Haldane phase\cite{Haldane}; beyond this point the ground-state corresponds to a pair of intertwined Haldane chains\cite{kolezhuk_connectivity,kolezhuk_prl,kolezhuk_prb}. The dimerized phase can be realized in spin-1 chain in the presence of biquadratic interaction $J_b({\bf S}_{i}\cdot {\bf S}_{i+1})^2$ or due to the three-site interaction $J_3 [({\bf S}_{i-1}\cdot {\bf S}_i)({\bf S}_{i}\cdot {\bf S}_{i+1})+\mathrm{h.c.}]$\cite{PhysRevB.40.4621,kulmper,XIAN1993437,PhysRevB.74.144426,michaud1,michaud2,PhysRevB.100.104426}. The latter is a generalization of the Majumdar-Ghosh point and realizes exact dimerized state at $J_3=J_1/[4S(S+1)-2]$ for any value of spin $S$\cite{michaud1}. In fact, for not too large next-nearest-neighbor coupling $J_2$, the exact dimerized state remains the ground-state along the line\cite{wang}: 
\begin{equation}
  \frac{J_3}{J_1-2J_2}=\frac{1}{4S(S+1)-2},
  \label{eq:exact}
\end{equation} 
making the $J_1-J_2-J_3$ model an ideal play-ground to study dimerization transitions in spin-S chains.  
The model is defined by the following microscopic Hamiltonian:
\begin{multline}
  H=J_1\sum_i {\bf S}_i\cdot{\bf S}_{i+1}+J_2\sum_i {\bf S}_{i-1}\cdot{\bf S}_{i+1}\\
  +J_3\sum_i\left[({\bf S}_{i-1}\cdot {\bf S}_i)({\bf S}_i\cdot {\bf S}_{i+1})+\mathrm{ h.c.}\right].
  \label{eq:j1j2j3s}
\end{multline}
where without loss of generality we fix $J_1=1$; in this paper we restrict ourselves to positive coupling constants $J_2,J_3\geq0$.

Numerical investigation of this model already lead to an impressive list of exotic quantum critical phenomena.
Among them a non-magnetic Ising phase transition between the fully-dimerized phase and the next-nearest neighbor (NNN-) Haldane phase in the spin-1 chain at which the  singlet-triplet gap remains open\cite{j1j2j3_short}; partially dimerized phases in spin-$3/2$ and $5/2$ chains\cite{spin_32paper,PhysRevB.105.174402} separated by Kosterlitz-Thouless transitions\cite{Kosterlitz} from the critical phases; and the emergence of magnetic floating phase - critical phase with incommensurate quasi-long-range correlations\cite{spin_32paper,PhysRevB.105.174402}. Another unusual feature revealed in these systems was a termination of the Wess-Zumino-Witten (WZW) SU(2)$_{2S}$ critical lines due to the presence of marginal operators in spin-1 and spin-$3/2$ chains. Above the end point the transition to the fully  dimerized phase is first order - a very counter-intuitive conclusion for a half-integer chain given that on one side of this transition the phase is critical\cite{affleck_haldane}.
 In both cases - in spin-1 and spin-3/2 chains - the marginal operators do not change the nature of the transition and the end point belongs to the same universality class as the critical line it terminates. 
 
The situation is radically different for spin chains with $S\geq5/2$ due to appearance of additional relevant operators\cite{AffleckGepner,PhysRevB.88.075132}. Realization of higher levels WZW SU(2)$_{k}$ universality classes is traditionally attributed to  integrable spin-$S$ chains with a microscopic Hamiltonian given by a polynomial of degree 2$S$ in $({\bf S_i}\cdot {\bf S_{i+1}})$\cite{kulish,takhtajan,babujian} that artificially fine-tunes all relevant operators to vanish. However, for WZW SU(2)$_k$ critical theory with $5\leq k< 10$ there are only two relevant operators consistent with the $\mathbb{Z}_2$ translation symmetry. This means that two independent parameters, as in the $J_1-J_2-J_3$ model, might be sufficient to fine-tune the system to the WZW SU(2)$_{2S}$ critical point. Recently, this has been reported in spin-$5/2$ chain where  SU(2)$_{5}$ critical points appear at the end of the extended SU(2)$_3$ critical line\cite{PhysRevB.105.174402}. First relevant operator is responsible for the appearance of the dimerized phase; the corresponding coupling constant changes its sign at the transition. The second relevant operator controls the type of this transition: when its coupling constant is positive, the transition is first order; when it is zero both operators vanish and the critical theory is fine-tuned to WZW SU(2)$_{5}$; when it is negative the underlying critical theory renormalizes to the lower level, SU(2)$_3$.  For small values of $J_2$ and $J_3$ when the coupling constant of both relevant operators are negative, the extended critical phase is described by the WZW SU(2)$_1$\cite{affleck_haldane} critical theory. According to the recent field theory prediction in the presence of both SU(2) and a discrete  $\mathbb{Z}_2$ symmetry a renormalization-group flow is only possible between WZW SU(2)$_k$ theories if the parity of the level index $k$ does not change\cite{PhysRevLett.118.021601}. The spectacular sequence of WZW criticalities  with  odd levels $k=1,3$ and $5$ reported in spin-$5/2$ chain confirms this theory prediction\cite{PhysRevB.105.174402}. At the same time possible renormalization between even levels remains unexplored numerically.  

In the present paper we aim to fill this gap by looking at the critical properties of the spin-3 $J_1-J_2-J_3$ chain.
In addition, we will explore the situation when there are several possible candidates with lower levels $k$ that satisfy the parity constraints.
 We will show that at the isolated point the transition is fine-tuned to the WZW SU(2)$_6$ that appears as an end point of WZW SU(2)$_4$ critical line.   We address this problem numerically with a state-of-the-art density matrix renormalization group (DMRG) algorithm\cite{dmrg1,dmrg2,dmrg3,dmrg4}. Throughout the paper, unless explicitly stated otherwise, we use a chain with an even number of sites and open boundary conditions. In two-site DMRG we typically keep up to 1500 states, perform 6 sweeps and discard singular values smaller than $10^{-8}$.

\section{Phase diagram}

\begin{figure}[t!]
\centering 
\includegraphics[width=0.5\textwidth]{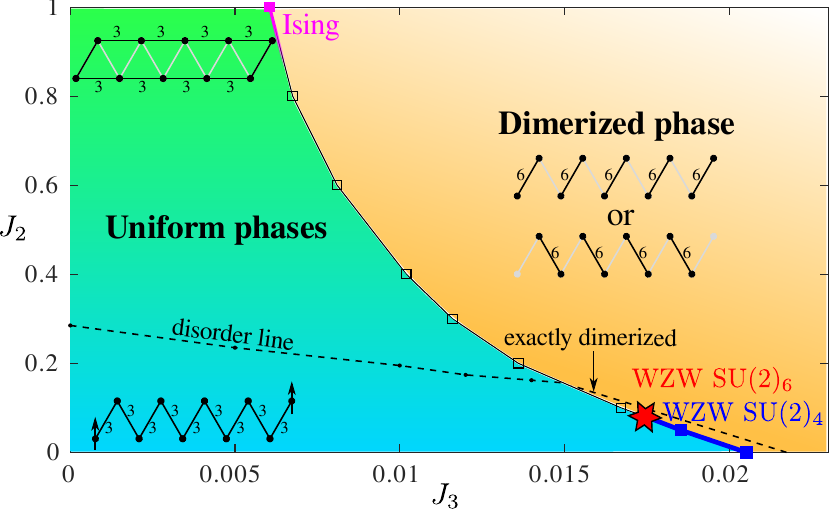}
\caption{Phase diagram of the $S=3$ chain with next-nearest-neighbor $J_2$ and three-site interactions $J_3$. There are at least two topologically distinct uniform phases;  according to Ref.\onlinecite{PhysRevB.100.014438} there might be four of them separated by topological transitions; their exact location are outside of the scope of this paper. Fully dimerized phase spontaneously breaks translation symmetry and has two-fold degenerate ground-states. The fully dimerized phase is separated from the (topologically non-trivial) uniform phase by continuous  WZW SU(2)$_4$ transition along the blue line that terminates at the end point (red star) located at $J_2\approx0.08$, beyond which the transition is first order. For large values of $J_2$ the transition between (topologically trivial) uniform phase and the fully-dimerized one is consistent with Ising universality class. The sketches are visualizations of the corresponding phases in terms of valence bond singlets (VBS): numbers near black lines state for the total number of VBS singlets at the corresponding bond. }
\label{fig:phasediag}
\end{figure}

We present a basic phase diagram of $J_1-J_2-J_3$ model for spin-3 chain obtained numerically in Fig.\ref{fig:phasediag}.
According to the Haldane's conjecture\cite{Haldane} integer-spin chain at $J_2=J_3=0$ is in the uniform and gapped phase. It is instructive to visualize this phase in terms of valence bond singlets (VBS) - distribution of effective spin-1/2 singlets. To match the total spin $S=3$ six VBS have to terminate at each lattice site. The phase at $J_2=J_3=0$ corresponds to a uniform VBS covering of the lattice with 3 spin-1/2 singlets per nearest-neighbor bond. In this representation it becomes obvious that there are three effective spins-1/2 at each end of the chain that remain unpaired and form spin-3/2 edge states, however only spin-1/2 edge states are topologically protected. At large $J_2$ by analogy with spin-1 case one might expect the VBS singlets to occupy next-nearest-neighbor bonds - the spin-3 analogue of the NNN-Haldane phase. This phase is topologically trivial and therefore the system undergoes at least one topological transition upon increasing $J_2$. In fact, Berry phase calculations on system sizes up to $N=10$ sites have shown that spin-3 $J_1-J_2$ chain undergoes a sequence of three topological transitions\cite{PhysRevB.100.014438}. Verification of whether all of these transitions will remain present in the thermodynamic limit and for finite values of $J_3$ is an extremely challenging numerical task that we leave for future investigations. But it is important to keep in mind that all these phases are uniform and show no sign of dimerization \footnote{In our finite-size calculations the dimerization measured in the middle of the chain in the uniflrm phases is $D_\mathrm{mid}\propto O(10^{-2})$. For comparison, the mid-chain dimerization in the partially dimerized phase in spin-5/2 chain is $D_\mathrm{mid}\propto O(1)$}.  From the Berry phase calculations we can expect the trivial uniform phase to be stable at least for $J_2\gtrsim 0.9$\cite{PhysRevB.100.014438}.  

For large $J_3$ the system is in the fully dimerized phase where every other nearest-neighbor bond is occupied by six VBS singlets. This results in two degenerate ground-states. The line of exact dimerization defined in Eq.\ref{eq:exact} terminates at $J_2\approx0.16$ - at this point the dimerization drops at once from $S(S+1)=12$ to zero signaling the presence of the first order transition, similar to the scenario reported in the spin-1 chain\cite{j1j2j3_short}. The transition between the topologically non-trivial uniform phase and the dimerized one is continuous for $J_2\lesssim 0.08$ in the WZW SU(2)$_4$ universality class, at $J_2\approx0.08$ this critical line terminates with the SU(2)$_6$ end point. At large $J_2$ the transition is between the two topologically trivial phases and is consistent with a non-magnetic Ising transition. In the next two sections we provide details on the nature of these continuous transitions.

\begin{figure}[t!]
\centering 
\includegraphics[width=0.5\textwidth]{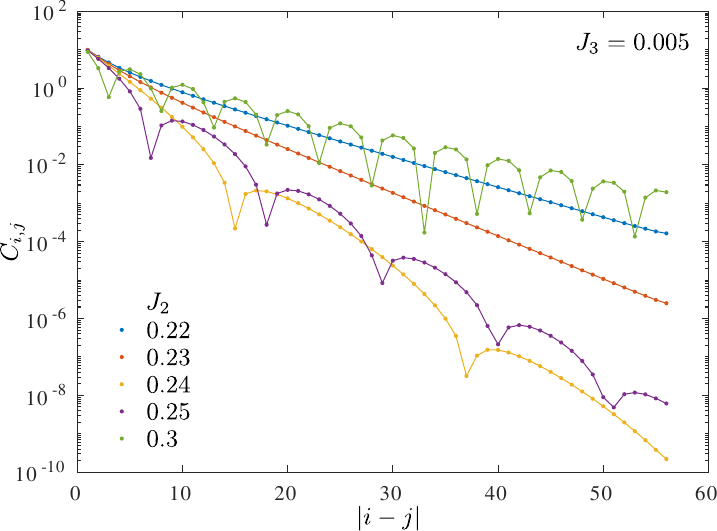}
\caption{Scaling of the connected correlation functions $C_{i,j}=\langle {\bf S}_i \cdot {\bf S}_j\rangle-\langle {\bf S}_i\rangle \cdot\langle {\bf S}_j\rangle$ with distance between spins for $J_3=0.005$ and various values of $J_3$. Starting from $J_2\approx 0.24$ system demonstrates the presence of incommensurate short-range correlations.  }
\label{fig:icexamples}
\end{figure}

Previous investigations of the $J_1-J_2-J_3$ model also revealed that a large portion of the phase diagram is characterized by incommensurate spin-spin correlations\cite{kolezhuk_prl,kolezhuk_prb,roth,j1j2j3_long,spin_32paper}. In spin-3 chain incommensurability appears only as a short-range order. Dotted line inside the uniform phases in Fig.\ref{fig:phasediag} marks the disorder line along which incommensurability develops. Examples of the connected spin-spin correlation function $C_{i,j}=\langle {\bf S}_i \cdot {\bf S}_j\rangle-\langle {\bf S}_i\rangle \cdot\langle {\bf S}_j\rangle$  on both sides of the disorder line are provided in Fig.\ref{fig:icexamples}. Inside the dimerized phase the disorder line coincides with the exact dimerized line.

\section{From WZW SU(2)$_6$ to SU(2)$_4$}

\begin{figure}[t!]
\centering 
\includegraphics[width=0.5\textwidth]{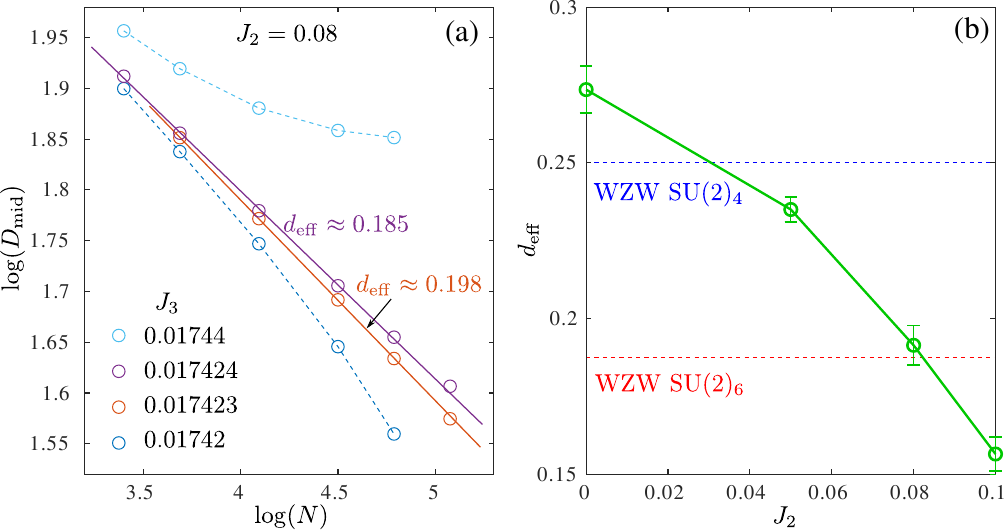}
\caption{ (a) Example of finite-size scaling of the middle-chain dimerization for $J_2=0.08$. The critical point is associated with the separatrix at $0.017423<J_3^c<0.017424$, and the slope gives an effective scaling dimension $0.185\lesssim d_\mathrm{eff}\lesssim0.198$. Dashed lines are guide to the eyes; solid lines are results of the fit. (b) Resulting value of the effective scaling dimension $d_\mathrm{eff}$ as a function of $J_2$ along the transition between the uniform and the fully dimerized phases. We associate the end point with the crossing points of the resulting curve and the horizontal line $d_{k=6}=3/16$ (dashed red). The dashed blue line stands for $d_{k=4}=1/4$.}
\label{fig:scaling}
\end{figure}

Let us now focus on the continuous transition at small values of $J_2$. In order locate the transition we look at the finite-size scaling of the middle-chain dimerization 
$D_\mathrm{mid}=|\langle {\bf S}_{N/2-1}\cdot {\bf S}_{N/2} \rangle-\langle {\bf S}_{N/2}\cdot {\bf S}_{N/2+1} \rangle|$.  In a log-log scale convex curves signals finite dimerization in the thermodynamic limit and therefore are associated with the dimerized phase.  In contrast, concave curves point towards the non-dimerized uniform phase. Quantum critical point between these two phases shows up as a separatrix on the finite-size scaling plot. By keeping track of the change of curvature we can narrow down the interval in which the quantum phase transition takes place. In Fig.\ref{fig:scaling}(a) we provide an example of finite-size scaling for $J_2=0.08$; the transition takes place between $J_3=0.017423$ where the scaling curves a bit downwards and $J_3=0.017424$ with data that slightly curve upwards. By fitting the main slope of these two finite-size scalings we estimate the upper and lower bounds of the critical scaling dimension $d$ for a given value of $J_2$. The  scaling dimensions extracted along the transition line are summarized in Fig.\ref{fig:scaling}(b). According to the conformal field theory (CFT), the scaling dimension of the WZW SU(2)$_{k}$ is given by $d=\frac{3}{2(2+k)}$, this implies $d_{k=4}=\frac{1}{4}$ and $d_{k=6}=\frac{3}{16}$. For small values of $J_2$ the scaling dimension $d$ is in a reasonable agreement with $d_{k=4}$,  we believe the discripancy between the exact value and the measured one is due to logarithmic corrections that appear when one of the relevant operators renormalizes to zero. It is very important to stress, that our results would not be consistent with another candidate satisfying Furuya and Oshikawa's parity rule\cite{PhysRevLett.118.021601} - the scaling dimension of  WZW SU(2)$_{2}$ is $d=\frac{3}{8}$ that significantly exceeds the scaling dimension extracted numerically.

\begin{figure}[t!]
\centering 
\includegraphics[width=0.4\textwidth]{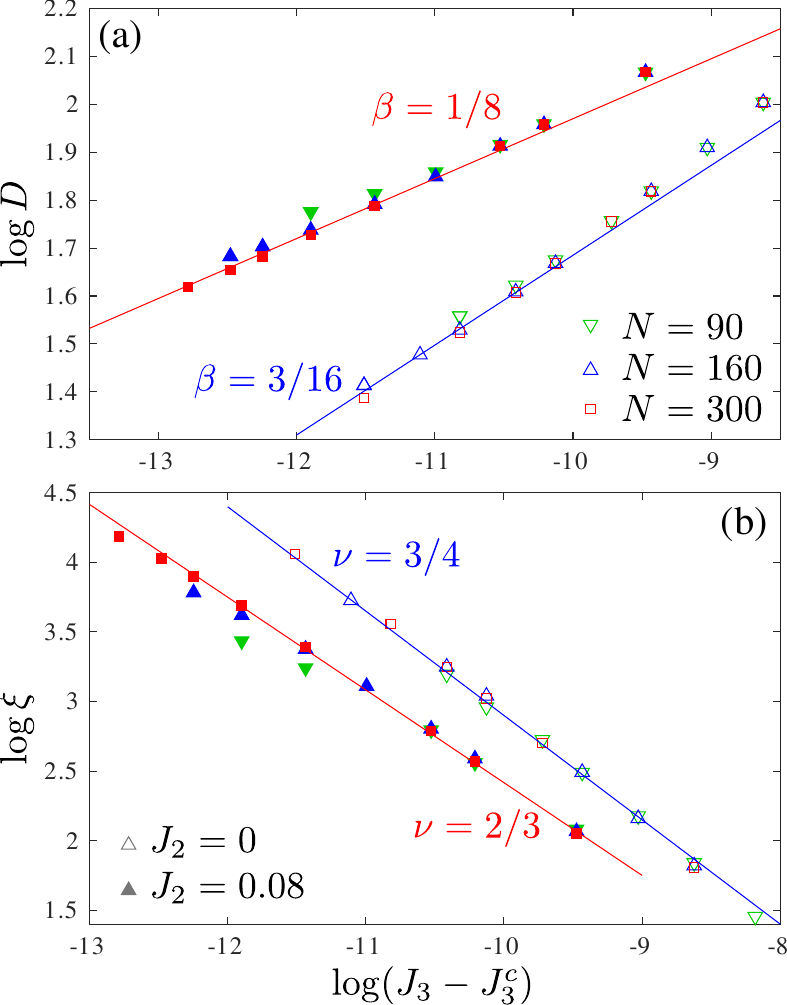}
\caption{ Scaling (a) of the middle chain dimerization $D$ and (b) of the correlation length $\xi$ upon approaching the phase transition from the fully dimerized phase to the uniform phase. Open symbols state for the horizontal cut at $J_2=0$, filled symbols for the horizontal cut through the end point at $J_2\approx 0.08$; the data are in excellent agreement with the CFT predictions for the critical exponents for WZW SU(2)$_4$ (blue lines) and SU(2)$_6$ (red lines).}
\label{fig:crit_exp}
\end{figure}

Following the procedure described in Ref.\onlinecite{PhysRevB.105.174402} we associate the point $J_2\approx0.08$ where the scaling dimension is consistent with  WZW SU(2)$_6$ and perform additional checks to confirm the universality class at the end points. First, we extract two critical exponents:  $\beta$ and  $\nu$ that measures the scaling of the order parameter (dimerization) and the correlation length with the distance to the transition. The results are presented in Fig.\ref{fig:crit_exp}. For both critical exponents we see an excellent agreement between numerical data (filled symbols) and CFT predictions for WZW SU(2)$_6$ $ \nu=\frac{2+k}{2k}$ and $\beta=\frac{3}{4k}$.  For comparison we also present results for $J_2=0$ that agree well with CFT predictions for SU(2)$_4$. Despite the fact that the two pairs of critical exponents are very close (0.125 vs 0.1875 for $\beta$ and 0.75 vs 0.667 for $\nu$) one can clearly distinguish these two cases in Fig.\ref{fig:crit_exp}.

 To the best of our knowledge the realization of the WZW SU(2)$_6$ phase transition in a non-integrable spin chain is reported for the first time and therefore desrves an additional check by extracting conformal the towers of states. Conformal tower is a very specific structure of excitation energy spectra that system takes at the conformal critical point and under specific cponformally-invariant boundary conditions\cite{diFrancesco}. Here we probe only the lowest level of each magnetization sector - the "envelop" of the conformal tower. The ground-state of the system with $N$-even is a singlet and thus has total spin $j=0$; a chain with odd number of sites has a ground-state with total spin $j=3$. We restrict ourselves to $S^z_\mathrm{tot}\leq12$. We construct the envelop of these two towers closely following Ref.\cite{AffleckGepner,PhysRevB.105.174402}.
 The predictions for the envelops are summarized in Table~\ref{tb:wzwsu2k6}.

\begin{table}[h!]
\centering
\begin{tabular}{|c|c|c|c|c|c|c|c|c|c|c|c|c|c|}
\multicolumn{7}{c}{j=0}\\
\hline 
$S^z_\mathrm{tot}$&0&1&2&3&4&5&6&7&8&9&10&11&12\\
\hline 
$(E-E_0)N/ \pi v$ &0&1&2&3&4&5&6&9&12&15&18&21&24\\
\hline 
\end{tabular}
\begin{tabular}{|c|c|c|c|c|c|c|c|c|c|c|c|}
\multicolumn{7}{c}{j=3}\\
\hline 
$S^z_\mathrm{tot}$&3&4&5&6&7&8&9&10&11&12\\
\hline 
$(E-E_0)N/ \pi v$, &0&2&4&6&8&10&12&16&20&24\\
\hline 
\end{tabular}
\caption{Lowest excitation energy with spin $S^z_\mathrm{tot}$ for both $j=0$ and $j=3$ WZW SU(2)$_{6}$ conformal towers.}
\label{tb:wzwsu2k6}
\end{table}

\begin{figure}[h!]
\centering 
\includegraphics[width=0.4\textwidth]{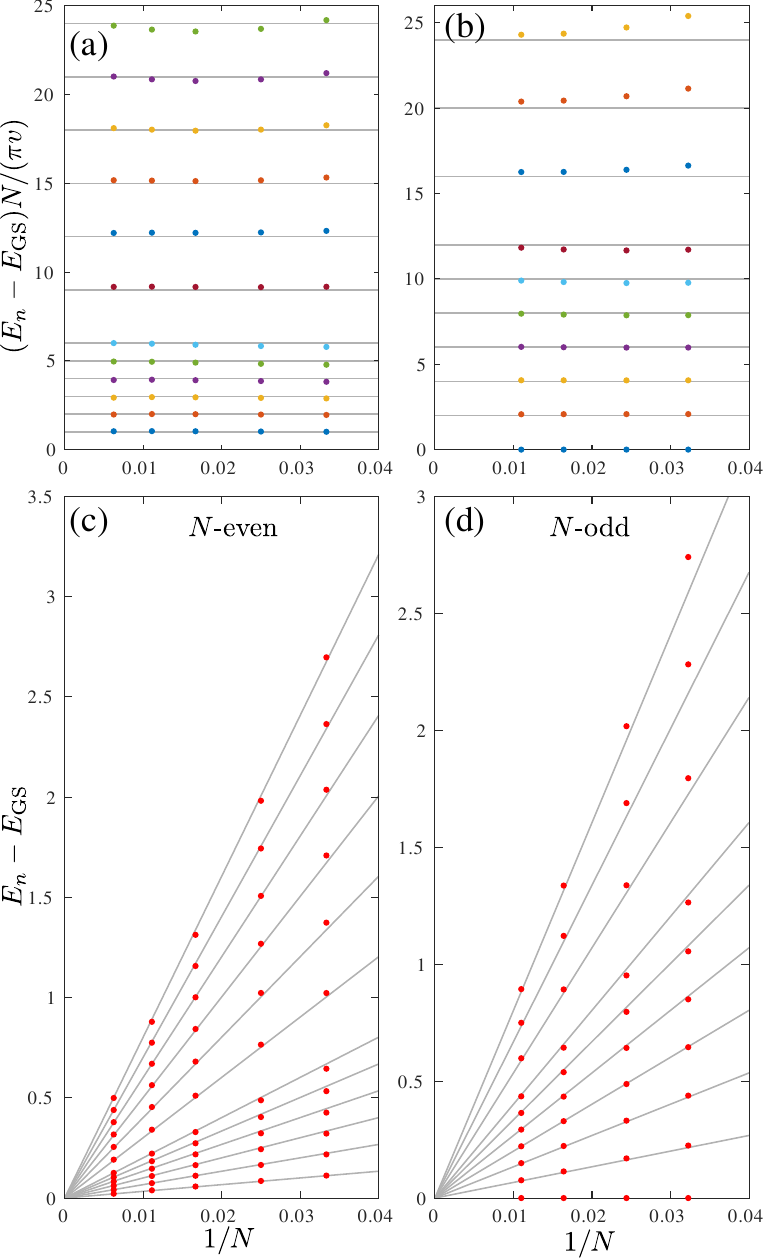}
\caption{ Conformal towers of states (top) and finite-size scaling of excitation spectrum (bottom) at the WZW SU(2)$_6$ end point. Symbols are DMRG data, lines are CFT predictions summarized in Table\ref{tb:wzwsu2k6}, sound velocity $v\approx 3.34\pm 0.05$ has been extracted by fitting the lowest two excitation energies with $N$-even.  For $N$ even, the ground state is in the singlet sector ($j=0$ tower), for $N$ odd, the ground state has total spin j=3. The towers show only the lowest state within each sector of total magnetization up to $S^z_\mathrm{tot}=12$.}
\label{fig:tower}
\end{figure}

Numerically we extract excitation energies for system sizes up to $N=160$ for $N$-even and up to $N=91$ for $N$-odd.
We extract a non-universal value of the sound velocity by fitting two lowest excitation energy levels for $N$-even; the extracted value is $v\approx 3.34\pm 0.05$. In Fig.\ref{fig:tower} we assume this value to draw CFT predictions without any further fitting or adjustments. The agreement with numerically extracted levels of the towers is excellent. 

\section{The second end of the SU(2)$_4$ line}

\begin{figure}[h!]
\centering 
\includegraphics[width=0.49\textwidth]{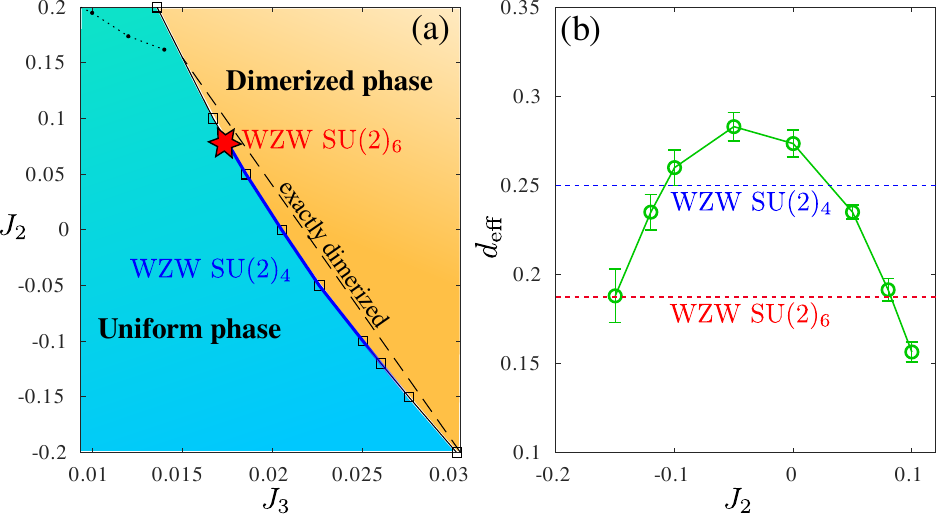}
\caption{ (a) Phase diagram of the spin-3 $J_1-J_2-J_3$ model zoomed around the transition to the dimerized phase for $-0.2< J_2 <0.2$. For negative $J_2$ the critical SU(2)$_4$ line terminates at $J_2\approx -0.12\pm 0.02$. Most likely the end point belongs to the same SU(2)$_4$ universality class as the critical line above it. (b)  Effective scaling dimension $d_\mathrm{eff}$  as a function of $J_2$ along the transition; this panel is an extended version of Fig.\ref{fig:scaling}(b). If the second SU(2)$_6$ end point exists it would take place at $J_2\approx-0.15$ where $d_\mathrm{eff}\approx d_{k=6}$. In Fig.\ref{fig:sep_tower}(b) we show the energy excitation spectrum at $J_2=-0.15$ that rules out this possibility. }
\label{fig:sep_pd}
\end{figure}

After identifying that  WZW SU(2)$_4$ critical line terminates with SU(2)$_6$ end point on one side, it is natural to wonder what happens at the second end of this line. 
 In Fig.\ref{fig:sep_pd}(a) we present the part of the phase diagram including the location of the dimerization transition for $J_2<0$. One can see that around $J_2\approx-0.2$ the transition line crosses the exactly dimerized line, implying that at least at this point the transition must be a first order. In practice, it means that the critical SU(2)$_4$ line has to terminate somewhere above. Based solely on the parity constraint for a level $k$ of the WZW criticality, there are three possibilities: {\it i)} the second end point belongs to SU(2)$_{k=6}$ universality class line the first one at $J_2\approx0.08$; {\it ii)} the end point is in the same ($k=4$) universality class as a critical line above it; and {\it iii)} the end point is critical with $k=2$.
 
\begin{figure}[h!]
\centering 
\includegraphics[width=0.4\textwidth]{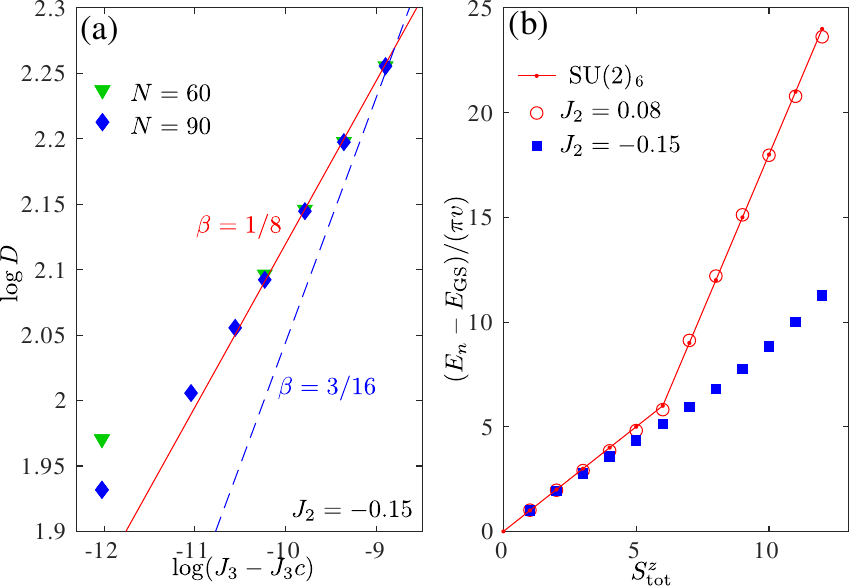}
\caption{ (a) Scaling of the middle chain dimerization $D$ upon approaching the phase transition from the fully dimerized phase to the uniform phase along $J_2=-0.15$. CFT predictions for the critical exponents $\beta=3/16$ for WZW SU(2)$_4$ and $\beta=1/8$ for  SU(2)$_6$ are show with the dashed blue and solid red lines. (b) Energy gap at $J_2=-0.15$ and $J_3^c\approx0.027564$ (blue squares) between the ground-state and the lowest energy state in the sector with total magnetization $S^z_\mathrm{tot}$ re-scaled with sound-velocity $v$. Results extracted at the end point at $J_2=0.8$ and $J_3^c\approx0.017423$ (red circles) and CFT predictions for SU(2)$_6$ (red dots) are shown for comparison.}
\label{fig:sep_tower}
\end{figure}

 Following similar procedure as in Fig.\ref{fig:sep_pd}(a) we extract the effective scaling dimension along the critical line also for $J_2<0$; these results are summarized in Fig.\ref{fig:sep_pd}(b). One can see that the scaling dimension $d_\mathrm{eff}$ never approaches $d=3/8$ of the WZW SU(2)$_{k=2}$ excluding the third scenario. Let us no take a closer look at the first option. If the second SU(2)$_{k=6}$ end point exists, it will be located at the point where $d_\mathrm{eff}\approx 3/16$. According to Fig.\ref{fig:sep_pd}(b) this happens around $J_2\approx-0.15$. In Fig.\ref{fig:sep_tower}(a) we present the scaling of the middle chain dimerization upon approaching this point along a horizontal cut in the dimerized phase which is in a reasonable agreement with  the critical exponent $\beta=1/8$ of the  WZW $k=6$ critical theory. However, from the field theory perspective, the presence of a marginal operator along SU(2)$_4$ line  that becomes relevant around  SU(2)$_6$ fixed point makes an emergence of two SU(2)$_6$ end points unlikely. Furthermore, as presented in Fig.\ref{fig:sep_tower}(b) the structure of the excitation spectrum extracted at $J_2=-0.15$  does not resemble the structure of the corresponding conformal tower for $k=6$.

Let us now consider the final option that, in fact, fits out numerical results: the critical SU(2)$_4$ line terminates at the end point in the same $k=4$ universality class. This can happens if, for instance, the coupling constant of the marginal operator of SU(2)$_4$ critical line changes its sign leading to a first order transition with slow opening of the gap below the end point. Similar physics has been reported for SU(2)$_2$ critical line in spin-1 chains\cite{j1j2j3_short}. This conjecture is supported by the excitation spectrum of Fig.\ref{fig:sep_tower}(b) that has a structure typical for the first order transitions with equally spaced levels due to almost non-interacting magnetic solitons\cite{j1j2j3_long,PhysRevB.101.115138}. When the density of solitons becomes too high the envelop starts to deviate from the linear slope. In the present case this happens beyond $S^z_\mathrm{tot}\approx 11$ on a chain with $N=40$ (see Fig.\ref{fig:sep_tower}(b)).
 Meanwhile, due to slow opening of the gap for some quantities including the order parameter the transition might look continuous due to finite-size effects. For $J_2<0$ numerical simulations are more challenging and we are limited to system sizes $N=90$, that might not be sufficient to resolve a weak first order transition at $J_2=-0.15$. To conclude, the most plausible scenario according to our numerical data is that SU(2)$_4$ critical line terminates around $J_2\approx 0.12\pm 0.02$ with the end point in the same universality class with $k=4$.

\section{Ising transition at large $J_2$}

We have shown so far that for small values of $J_2$ the transition to the dimerized phase is continuous and terminates at the end point at $J_2\approx0.08$. Beyond this point the transition is first order. This conclusion is supported by the observation that at the point where exact dimerized line hits the transition the order parameter jumps discontinuously. Now, we are going to ask ourselves whether this first order transition will eventually turn (again) into a continuous transition at large values of $J_2$. As a disclaimer, let us mention here that we leave  detailed investigation of the uniform phase(s) and phase transition(s) in the range $0.2\lesssim J_2\lesssim 0.8$ outside of the scope of this work. However, we want to share with our readers one very interesting observation.

\begin{figure}[h!]
\centering 
\includegraphics[width=0.4\textwidth]{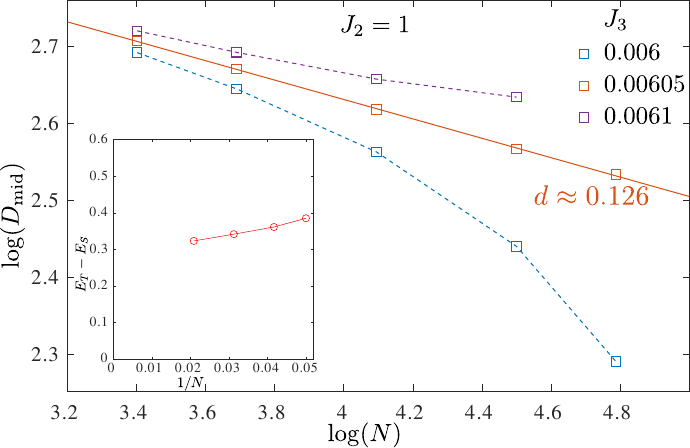}
\caption{ Finite-size scaling of the middle-chain dimerization for $J_2=1$. Separatrix is associated with the phase transition; the slope gives the effective scaling dimension that is in the excellent agreement with the CFT prediction $d=1/8$. Inset: Finite-size scaling of the singlet-triplet gap at the critical point $J_3^c\approx0.00605$. The magnetic gap remains open at the transition.  }
\label{fig:ising}
\end{figure}

In the region where the uniform phase is expected to be topologically trivial, the domain wall between the uniform and the dimerized phases are non-magnetic. This transition is therefore expected to take place entirely in the singlet sector. 
Since $\mathbb{Z}_2$ symmetry is broken at the transition it is natural to expect the underlying critical theory to be Ising CFT with scaling dimension $d=1/8$.
Our numerical results for the finite-size scaling of the order parameter presented in Fig.\ref{fig:ising} is in excellent agreement with this prediction. Non-magnetic nature of this transition is supported by finite-size scaling of the singlet-triplet gap presented in the inset of Fig.\ref{fig:ising} that shows no tendency to close in the thermodynamic limit.

\section{Discussion}
\label{sec:discussion}

To summarize, in the present paper we have studied dimerization transitions in a frustrated spin-3 chain.
We identified the end point in the WZW SU(2)$_6$ universality class into which the $J_1-J_2-J_3$ model with two independent parameters is fine tuned. The universality class of the underlying critical theory has been verified with the scaling dimension $d$ and conformal towers of states computed at the transitions as well as with the critical exponents $\beta$ and $\nu$ controlling the scaling of the order parameter and the divergence of the correlation length upon approaching the transition.

Away from the fine-tuned end point we detect that the critical theory due to the presence of second relevant operator re-normalizes to a critical theory of the lower level of the same parity - SU(2)$_4$. Eventually, the critical SU(2)$_4$ line terminates via the end point in the same universality class.  
This picture is different from the one reported for WZW transitions with odd $k$-levels, where SU(2)$_3$ line was terminated {\it on both sides} with the SU(2)$_5$ critical points. We belief that the difference is attributed to the presence of an additional marginal operator along the SU(2)$_4$ line. We hope that our results will stimulate further theoretical exploration of this problem.

  We see no signature of possible renormalization to the level  $k=2$. This lead us to a hypothesis that the renormalization group flow will lead towards the highest level with the right number of relevant operators. In other words, for frustrated spin model with $5/2\leq S< 5$ and two independent parameters one might expect SU(2)$_{2S}$ to be realized as the end points  terminating the WZW SU(2)$_4$ critical line  for integer- and  the SU(2)$_3$ line for half-integer spin chains - both independent on the value of spin $S$. The case of spin $S=5$ is less obvious due to the appearance of an additional marginal operator. Beyond that for $S>5$ a generic two-dimensional parameter space would  not suffice to fine-tune the system into WZW SU(2)$_{2S}$ critical points and we expect the critical lines for integer (half-integer) chains to be in SU(2)$_4$ (SU(2)$_3$) terminated  with SU(2)$_{10}$ (SU(2)$_9$) end points.
This hypothesis can be checked numerically with $S>3$, though, of course, computationally this is very challenging. One has to admit a possibility that for a given model, e.g. for $J_1-J_2-J_3$, the interval of continuous transition might shrink for large $S$ to the point when it can eventually disappear.

In addition, for large value of $J_2$ we report an appearance of non-magnetic Ising transition, previously detected in frustrated spin-1 chain. This supports previous conjecture that the appearance of the Ising transition at the boundary of the dimerized phase and large $J_2$ is generic for any integer spin chains.

\section{Acknowledgments} 
I am indebted to Frederic Mila and Ian Affleck for many inspiring discussions on dimerization transitions. I would like to thank Shunsuke Furuya and the anonymous referee for an inspiration to look at the second end point. This research has been supported by Delft Technology Fellowship. Numerical simulations have been performed with the Dutch national e-infrastructure with the support of the SURF Cooperative and  at the DelftBlue HPC.
\bibliographystyle{apsrev4-2}
\bibliography{bibliography}

\begin{thebibliography}{35}%
\makeatletter
\providecommand \@ifxundefined [1]{%
 \@ifx{#1\undefined}
}%
\providecommand \@ifnum [1]{%
 \ifnum #1\expandafter \@firstoftwo
 \else \expandafter \@secondoftwo
 \fi
}%
\providecommand \@ifx [1]{%
 \ifx #1\expandafter \@firstoftwo
 \else \expandafter \@secondoftwo
 \fi
}%
\providecommand \natexlab [1]{#1}%
\providecommand \enquote  [1]{``#1''}%
\providecommand \bibnamefont  [1]{#1}%
\providecommand \bibfnamefont [1]{#1}%
\providecommand \citenamefont [1]{#1}%
\providecommand \href@noop [0]{\@secondoftwo}%
\providecommand \href [0]{\begingroup \@sanitize@url \@href}%
\providecommand \@href[1]{\@@startlink{#1}\@@href}%
\providecommand \@@href[1]{\endgroup#1\@@endlink}%
\providecommand \@sanitize@url [0]{\catcode `\\12\catcode `\$12\catcode
  `\&12\catcode `\#12\catcode `\^12\catcode `\_12\catcode `\%12\relax}%
\providecommand \@@startlink[1]{}%
\providecommand \@@endlink[0]{}%
\providecommand \url  [0]{\begingroup\@sanitize@url \@url }%
\providecommand \@url [1]{\endgroup\@href {#1}{\urlprefix }}%
\providecommand \urlprefix  [0]{URL }%
\providecommand \Eprint [0]{\href }%
\providecommand \doibase [0]{https://doi.org/}%
\providecommand \selectlanguage [0]{\@gobble}%
\providecommand \bibinfo  [0]{\@secondoftwo}%
\providecommand \bibfield  [0]{\@secondoftwo}%
\providecommand \translation [1]{[#1]}%
\providecommand \BibitemOpen [0]{}%
\providecommand \bibitemStop [0]{}%
\providecommand \bibitemNoStop [0]{.\EOS\space}%
\providecommand \EOS [0]{\spacefactor3000\relax}%
\providecommand \BibitemShut  [1]{\csname bibitem#1\endcsname}%
\let\auto@bib@innerbib\@empty
\bibitem [{\citenamefont {Kosterlitz}\ and\ \citenamefont
  {Thouless}(1973)}]{Kosterlitz}%
  \BibitemOpen
  \bibfield  {author} {\bibinfo {author} {\bibfnamefont {J.~M.}\ \bibnamefont
  {Kosterlitz}}\ and\ \bibinfo {author} {\bibfnamefont {D.~J.}\ \bibnamefont
  {Thouless}},\ }\href {http://stacks.iop.org/0022-3719/6/i=7/a=010} {\bibfield
   {journal} {\bibinfo  {journal} {Journal of Physics C: Solid State Physics}\
  }\textbf {\bibinfo {volume} {6}},\ \bibinfo {pages} {1181} (\bibinfo {year}
  {1973})}\BibitemShut {NoStop}%
\bibitem [{\citenamefont {Majumdar}\ and\ \citenamefont
  {Ghosh}(1969)}]{MajumdarGhosh}%
  \BibitemOpen
  \bibfield  {author} {\bibinfo {author} {\bibfnamefont {C.~K.}\ \bibnamefont
  {Majumdar}}\ and\ \bibinfo {author} {\bibfnamefont {D.~K.}\ \bibnamefont
  {Ghosh}},\ }\href {https://doi.org/http://dx.doi.org/10.1063/1.1664978}
  {\bibfield  {journal} {\bibinfo  {journal} {Journal of Mathematical Physics}\
  }\textbf {\bibinfo {volume} {10}},\ \bibinfo {pages} {1388} (\bibinfo {year}
  {1969})}\BibitemShut {NoStop}%
\bibitem [{\citenamefont {Okamoto}\ and\ \citenamefont
  {Nomura}(1992)}]{okamoto}%
  \BibitemOpen
  \bibfield  {author} {\bibinfo {author} {\bibfnamefont {K.}~\bibnamefont
  {Okamoto}}\ and\ \bibinfo {author} {\bibfnamefont {K.}~\bibnamefont
  {Nomura}},\ }\href
  {https://doi.org/http://dx.doi.org/10.1016/0375-9601(92)90823-5} {\bibfield
  {journal} {\bibinfo  {journal} {Physics Letters A}\ }\textbf {\bibinfo
  {volume} {169}},\ \bibinfo {pages} {433 } (\bibinfo {year}
  {1992})}\BibitemShut {NoStop}%
\bibitem [{\citenamefont {Haldane}(1983)}]{Haldane}%
  \BibitemOpen
  \bibfield  {author} {\bibinfo {author} {\bibfnamefont {F.~D.~M.}\
  \bibnamefont {Haldane}},\ }\href {https://doi.org/16/0375-9601(83)90631-X}
  {\bibfield  {journal} {\bibinfo  {journal} {Physics Letters A}\ }\textbf
  {\bibinfo {volume} {93}},\ \bibinfo {pages} {464} (\bibinfo {year}
  {1983})}\BibitemShut {NoStop}%
\bibitem [{\citenamefont {Kolezhuk}\ and\ \citenamefont
  {Schollw\"ock}(2002)}]{kolezhuk_connectivity}%
  \BibitemOpen
  \bibfield  {author} {\bibinfo {author} {\bibfnamefont {A.~K.}\ \bibnamefont
  {Kolezhuk}}\ and\ \bibinfo {author} {\bibfnamefont {U.}~\bibnamefont
  {Schollw\"ock}},\ }\href {https://doi.org/10.1103/PhysRevB.65.100401}
  {\bibfield  {journal} {\bibinfo  {journal} {Phys. Rev. B}\ }\textbf {\bibinfo
  {volume} {65}},\ \bibinfo {pages} {100401} (\bibinfo {year}
  {2002})}\BibitemShut {NoStop}%
\bibitem [{\citenamefont {Kolezhuk}\ \emph {et~al.}(1996)\citenamefont
  {Kolezhuk}, \citenamefont {Roth},\ and\ \citenamefont
  {Schollw\"ock}}]{kolezhuk_prl}%
  \BibitemOpen
  \bibfield  {author} {\bibinfo {author} {\bibfnamefont {A.}~\bibnamefont
  {Kolezhuk}}, \bibinfo {author} {\bibfnamefont {R.}~\bibnamefont {Roth}},\
  and\ \bibinfo {author} {\bibfnamefont {U.}~\bibnamefont {Schollw\"ock}},\
  }\href {https://doi.org/10.1103/PhysRevLett.77.5142} {\bibfield  {journal}
  {\bibinfo  {journal} {Phys. Rev. Lett.}\ }\textbf {\bibinfo {volume} {77}},\
  \bibinfo {pages} {5142} (\bibinfo {year} {1996})}\BibitemShut {NoStop}%
\bibitem [{\citenamefont {Kolezhuk}\ \emph {et~al.}(1997)\citenamefont
  {Kolezhuk}, \citenamefont {Roth},\ and\ \citenamefont
  {Schollw\"ock}}]{kolezhuk_prb}%
  \BibitemOpen
  \bibfield  {author} {\bibinfo {author} {\bibfnamefont {A.}~\bibnamefont
  {Kolezhuk}}, \bibinfo {author} {\bibfnamefont {R.}~\bibnamefont {Roth}},\
  and\ \bibinfo {author} {\bibfnamefont {U.}~\bibnamefont {Schollw\"ock}},\
  }\href {https://doi.org/10.1103/PhysRevB.55.8928} {\bibfield  {journal}
  {\bibinfo  {journal} {Phys. Rev. B}\ }\textbf {\bibinfo {volume} {55}},\
  \bibinfo {pages} {8928} (\bibinfo {year} {1997})}\BibitemShut {NoStop}%
\bibitem [{\citenamefont {Barber}\ and\ \citenamefont
  {Batchelor}(1989)}]{PhysRevB.40.4621}%
  \BibitemOpen
  \bibfield  {author} {\bibinfo {author} {\bibfnamefont {M.~N.}\ \bibnamefont
  {Barber}}\ and\ \bibinfo {author} {\bibfnamefont {M.~T.}\ \bibnamefont
  {Batchelor}},\ }\href {https://doi.org/10.1103/PhysRevB.40.4621} {\bibfield
  {journal} {\bibinfo  {journal} {Phys. Rev. B}\ }\textbf {\bibinfo {volume}
  {40}},\ \bibinfo {pages} {4621} (\bibinfo {year} {1989})}\BibitemShut
  {NoStop}%
\bibitem [{\citenamefont {Klümper}(1989)}]{kulmper}%
  \BibitemOpen
  \bibfield  {author} {\bibinfo {author} {\bibfnamefont {A.}~\bibnamefont
  {Klümper}},\ }\href {https://doi.org/10.1209/0295-5075/9/8/013} {\bibfield
  {journal} {\bibinfo  {journal} {Europhysics Letters}\ }\textbf {\bibinfo
  {volume} {9}},\ \bibinfo {pages} {815} (\bibinfo {year} {1989})}\BibitemShut
  {NoStop}%
\bibitem [{\citenamefont {Xian}(1993)}]{XIAN1993437}%
  \BibitemOpen
  \bibfield  {author} {\bibinfo {author} {\bibfnamefont {Y.}~\bibnamefont
  {Xian}},\ }\href
  {https://doi.org/https://doi.org/10.1016/0375-9601(93)90602-V} {\bibfield
  {journal} {\bibinfo  {journal} {Physics Letters A}\ }\textbf {\bibinfo
  {volume} {183}},\ \bibinfo {pages} {437} (\bibinfo {year}
  {1993})}\BibitemShut {NoStop}%
\bibitem [{\citenamefont {L\"auchli}\ \emph {et~al.}(2006)\citenamefont
  {L\"auchli}, \citenamefont {Schmid},\ and\ \citenamefont
  {Trebst}}]{PhysRevB.74.144426}%
  \BibitemOpen
  \bibfield  {author} {\bibinfo {author} {\bibfnamefont {A.}~\bibnamefont
  {L\"auchli}}, \bibinfo {author} {\bibfnamefont {G.}~\bibnamefont {Schmid}},\
  and\ \bibinfo {author} {\bibfnamefont {S.}~\bibnamefont {Trebst}},\ }\href
  {https://doi.org/10.1103/PhysRevB.74.144426} {\bibfield  {journal} {\bibinfo
  {journal} {Phys. Rev. B}\ }\textbf {\bibinfo {volume} {74}},\ \bibinfo
  {pages} {144426} (\bibinfo {year} {2006})}\BibitemShut {NoStop}%
\bibitem [{\citenamefont {Michaud}\ \emph {et~al.}(2012)\citenamefont
  {Michaud}, \citenamefont {Vernay}, \citenamefont {Manmana},\ and\
  \citenamefont {Mila}}]{michaud1}%
  \BibitemOpen
  \bibfield  {author} {\bibinfo {author} {\bibfnamefont {F.}~\bibnamefont
  {Michaud}}, \bibinfo {author} {\bibfnamefont {F.}~\bibnamefont {Vernay}},
  \bibinfo {author} {\bibfnamefont {S.~R.}\ \bibnamefont {Manmana}},\ and\
  \bibinfo {author} {\bibfnamefont {F.}~\bibnamefont {Mila}},\ }\href
  {https://doi.org/10.1103/PhysRevLett.108.127202} {\bibfield  {journal}
  {\bibinfo  {journal} {Phys. Rev. Lett.}\ }\textbf {\bibinfo {volume} {108}},\
  \bibinfo {pages} {127202} (\bibinfo {year} {2012})}\BibitemShut {NoStop}%
\bibitem [{\citenamefont {Michaud}\ \emph {et~al.}(2013)\citenamefont
  {Michaud}, \citenamefont {Manmana},\ and\ \citenamefont {Mila}}]{michaud2}%
  \BibitemOpen
  \bibfield  {author} {\bibinfo {author} {\bibfnamefont {F.}~\bibnamefont
  {Michaud}}, \bibinfo {author} {\bibfnamefont {S.~R.}\ \bibnamefont
  {Manmana}},\ and\ \bibinfo {author} {\bibfnamefont {F.}~\bibnamefont
  {Mila}},\ }\href {https://doi.org/10.1103/PhysRevB.87.140404} {\bibfield
  {journal} {\bibinfo  {journal} {Phys. Rev. B}\ }\textbf {\bibinfo {volume}
  {87}},\ \bibinfo {pages} {140404} (\bibinfo {year} {2013})}\BibitemShut
  {NoStop}%
\bibitem [{\citenamefont {Chepiga}\ and\ \citenamefont
  {Mila}(2019)}]{PhysRevB.100.104426}%
  \BibitemOpen
  \bibfield  {author} {\bibinfo {author} {\bibfnamefont {N.}~\bibnamefont
  {Chepiga}}\ and\ \bibinfo {author} {\bibfnamefont {F.}~\bibnamefont {Mila}},\
  }\href {https://doi.org/10.1103/PhysRevB.100.104426} {\bibfield  {journal}
  {\bibinfo  {journal} {Phys. Rev. B}\ }\textbf {\bibinfo {volume} {100}},\
  \bibinfo {pages} {104426} (\bibinfo {year} {2019})}\BibitemShut {NoStop}%
\bibitem [{\citenamefont {Wang}\ \emph {et~al.}(2013)\citenamefont {Wang},
  \citenamefont {Furuya}, \citenamefont {Nakamura},\ and\ \citenamefont
  {Komakura}}]{wang}%
  \BibitemOpen
  \bibfield  {author} {\bibinfo {author} {\bibfnamefont {Z.-Y.}\ \bibnamefont
  {Wang}}, \bibinfo {author} {\bibfnamefont {S.~C.}\ \bibnamefont {Furuya}},
  \bibinfo {author} {\bibfnamefont {M.}~\bibnamefont {Nakamura}},\ and\
  \bibinfo {author} {\bibfnamefont {R.}~\bibnamefont {Komakura}},\ }\href
  {https://doi.org/10.1103/PhysRevB.88.224419} {\bibfield  {journal} {\bibinfo
  {journal} {Phys. Rev. B}\ }\textbf {\bibinfo {volume} {88}},\ \bibinfo
  {pages} {224419} (\bibinfo {year} {2013})}\BibitemShut {NoStop}%
\bibitem [{\citenamefont {Chepiga}\ \emph
  {et~al.}(2016{\natexlab{a}})\citenamefont {Chepiga}, \citenamefont
  {Affleck},\ and\ \citenamefont {Mila}}]{j1j2j3_short}%
  \BibitemOpen
  \bibfield  {author} {\bibinfo {author} {\bibfnamefont {N.}~\bibnamefont
  {Chepiga}}, \bibinfo {author} {\bibfnamefont {I.}~\bibnamefont {Affleck}},\
  and\ \bibinfo {author} {\bibfnamefont {F.}~\bibnamefont {Mila}},\ }\href
  {https://doi.org/10.1103/PhysRevB.93.241108} {\bibfield  {journal} {\bibinfo
  {journal} {Phys. Rev. B}\ }\textbf {\bibinfo {volume} {93}},\ \bibinfo
  {pages} {241108} (\bibinfo {year} {2016}{\natexlab{a}})}\BibitemShut
  {NoStop}%
\bibitem [{\citenamefont {Chepiga}\ \emph {et~al.}(2020)\citenamefont
  {Chepiga}, \citenamefont {Affleck},\ and\ \citenamefont
  {Mila}}]{spin_32paper}%
  \BibitemOpen
  \bibfield  {author} {\bibinfo {author} {\bibfnamefont {N.}~\bibnamefont
  {Chepiga}}, \bibinfo {author} {\bibfnamefont {I.}~\bibnamefont {Affleck}},\
  and\ \bibinfo {author} {\bibfnamefont {F.}~\bibnamefont {Mila}},\ }\href
  {https://doi.org/10.1103/PhysRevB.101.174407} {\bibfield  {journal} {\bibinfo
   {journal} {Phys. Rev. B}\ }\textbf {\bibinfo {volume} {101}},\ \bibinfo
  {pages} {174407} (\bibinfo {year} {2020})}\BibitemShut {NoStop}%
\bibitem [{\citenamefont {Chepiga}\ \emph {et~al.}(2022)\citenamefont
  {Chepiga}, \citenamefont {Affleck},\ and\ \citenamefont
  {Mila}}]{PhysRevB.105.174402}%
  \BibitemOpen
  \bibfield  {author} {\bibinfo {author} {\bibfnamefont {N.}~\bibnamefont
  {Chepiga}}, \bibinfo {author} {\bibfnamefont {I.}~\bibnamefont {Affleck}},\
  and\ \bibinfo {author} {\bibfnamefont {F.}~\bibnamefont {Mila}},\ }\href
  {https://doi.org/10.1103/PhysRevB.105.174402} {\bibfield  {journal} {\bibinfo
   {journal} {Phys. Rev. B}\ }\textbf {\bibinfo {volume} {105}},\ \bibinfo
  {pages} {174402} (\bibinfo {year} {2022})}\BibitemShut {NoStop}%
\bibitem [{\citenamefont {Affleck}\ and\ \citenamefont
  {Haldane}(1987)}]{affleck_haldane}%
  \BibitemOpen
  \bibfield  {author} {\bibinfo {author} {\bibfnamefont {I.}~\bibnamefont
  {Affleck}}\ and\ \bibinfo {author} {\bibfnamefont {F.~D.~M.}\ \bibnamefont
  {Haldane}},\ }\href {https://doi.org/10.1103/PhysRevB.36.5291} {\bibfield
  {journal} {\bibinfo  {journal} {Phys. Rev. B}\ }\textbf {\bibinfo {volume}
  {36}},\ \bibinfo {pages} {5291} (\bibinfo {year} {1987})}\BibitemShut
  {NoStop}%
\bibitem [{\citenamefont {Affleck}\ \emph {et~al.}(1990)\citenamefont
  {Affleck}, \citenamefont {Gepner}, \citenamefont {Schulz},\ and\
  \citenamefont {Ziman}}]{AffleckGepner}%
  \BibitemOpen
  \bibfield  {author} {\bibinfo {author} {\bibfnamefont {I.}~\bibnamefont
  {Affleck}}, \bibinfo {author} {\bibfnamefont {D.}~\bibnamefont {Gepner}},
  \bibinfo {author} {\bibfnamefont {H.~J.}\ \bibnamefont {Schulz}},\ and\
  \bibinfo {author} {\bibfnamefont {T.}~\bibnamefont {Ziman}},\ }\href
  {http://stacks.iop.org/0305-4470/23/i=20/a=534} {\bibfield  {journal}
  {\bibinfo  {journal} {Journal of Physics A: Mathematical and General}\
  }\textbf {\bibinfo {volume} {23}},\ \bibinfo {pages} {4725} (\bibinfo {year}
  {1990})}\BibitemShut {NoStop}%
\bibitem [{\citenamefont {Capponi}\ \emph {et~al.}(2013)\citenamefont
  {Capponi}, \citenamefont {Lecheminant},\ and\ \citenamefont
  {Moliner}}]{PhysRevB.88.075132}%
  \BibitemOpen
  \bibfield  {author} {\bibinfo {author} {\bibfnamefont {S.}~\bibnamefont
  {Capponi}}, \bibinfo {author} {\bibfnamefont {P.}~\bibnamefont
  {Lecheminant}},\ and\ \bibinfo {author} {\bibfnamefont {M.}~\bibnamefont
  {Moliner}},\ }\href {https://doi.org/10.1103/PhysRevB.88.075132} {\bibfield
  {journal} {\bibinfo  {journal} {Phys. Rev. B}\ }\textbf {\bibinfo {volume}
  {88}},\ \bibinfo {pages} {075132} (\bibinfo {year} {2013})}\BibitemShut
  {NoStop}%
\bibitem [{\citenamefont {{Kulish}}\ \emph {et~al.}(1981)\citenamefont
  {{Kulish}}, \citenamefont {{Reshetikhin}},\ and\ \citenamefont
  {{Sklyanin}}}]{kulish}%
  \BibitemOpen
  \bibfield  {author} {\bibinfo {author} {\bibfnamefont {P.~P.}\ \bibnamefont
  {{Kulish}}}, \bibinfo {author} {\bibfnamefont {N.~Y.}\ \bibnamefont
  {{Reshetikhin}}},\ and\ \bibinfo {author} {\bibfnamefont {E.~K.}\
  \bibnamefont {{Sklyanin}}},\ }\href {https://doi.org/10.1007/BF02285311}
  {\bibfield  {journal} {\bibinfo  {journal} {Letters in Mathematical Physics}\
  }\textbf {\bibinfo {volume} {5}},\ \bibinfo {pages} {393} (\bibinfo {year}
  {1981})}\BibitemShut {NoStop}%
\bibitem [{\citenamefont {Takhtajan}(1982)}]{takhtajan}%
  \BibitemOpen
  \bibfield  {author} {\bibinfo {author} {\bibfnamefont {L.~A.}\ \bibnamefont
  {Takhtajan}},\ }\href {https://doi.org/DOI: 10.1016/0375-9601(82)90764-2}
  {\bibfield  {journal} {\bibinfo  {journal} {Physics Letters A}\ }\textbf
  {\bibinfo {volume} {87}},\ \bibinfo {pages} {479 } (\bibinfo {year}
  {1982})}\BibitemShut {NoStop}%
\bibitem [{\citenamefont {Babujian}(1983)}]{babujian}%
  \BibitemOpen
  \bibfield  {author} {\bibinfo {author} {\bibfnamefont {H.~M.}\ \bibnamefont
  {Babujian}},\ }\href {https://doi.org/DOI: 10.1016/0550-3213(83)90668-5}
  {\bibfield  {journal} {\bibinfo  {journal} {Nuclear Physics B}\ }\textbf
  {\bibinfo {volume} {215}},\ \bibinfo {pages} {317 } (\bibinfo {year}
  {1983})}\BibitemShut {NoStop}%
\bibitem [{\citenamefont {Furuya}\ and\ \citenamefont
  {Oshikawa}(2017)}]{PhysRevLett.118.021601}%
  \BibitemOpen
  \bibfield  {author} {\bibinfo {author} {\bibfnamefont {S.~C.}\ \bibnamefont
  {Furuya}}\ and\ \bibinfo {author} {\bibfnamefont {M.}~\bibnamefont
  {Oshikawa}},\ }\href {https://doi.org/10.1103/PhysRevLett.118.021601}
  {\bibfield  {journal} {\bibinfo  {journal} {Phys. Rev. Lett.}\ }\textbf
  {\bibinfo {volume} {118}},\ \bibinfo {pages} {021601} (\bibinfo {year}
  {2017})}\BibitemShut {NoStop}%
\bibitem [{\citenamefont {White}(1992)}]{dmrg1}%
  \BibitemOpen
  \bibfield  {author} {\bibinfo {author} {\bibfnamefont {S.~R.}\ \bibnamefont
  {White}},\ }\href {https://doi.org/10.1103/PhysRevLett.69.2863} {\bibfield
  {journal} {\bibinfo  {journal} {Phys. Rev. Lett.}\ }\textbf {\bibinfo
  {volume} {69}},\ \bibinfo {pages} {2863} (\bibinfo {year}
  {1992})}\BibitemShut {NoStop}%
\bibitem [{\citenamefont {Schollw\"ock}(2005)}]{dmrg2}%
  \BibitemOpen
  \bibfield  {author} {\bibinfo {author} {\bibfnamefont {U.}~\bibnamefont
  {Schollw\"ock}},\ }\href {https://doi.org/10.1103/RevModPhys.77.259}
  {\bibfield  {journal} {\bibinfo  {journal} {Rev. Mod. Phys.}\ }\textbf
  {\bibinfo {volume} {77}},\ \bibinfo {pages} {259} (\bibinfo {year}
  {2005})}\BibitemShut {NoStop}%
\bibitem [{\citenamefont {\"Ostlund}\ and\ \citenamefont
  {Rommer}(1995)}]{dmrg3}%
  \BibitemOpen
  \bibfield  {author} {\bibinfo {author} {\bibfnamefont {S.}~\bibnamefont
  {\"Ostlund}}\ and\ \bibinfo {author} {\bibfnamefont {S.}~\bibnamefont
  {Rommer}},\ }\href {https://doi.org/10.1103/PhysRevLett.75.3537} {\bibfield
  {journal} {\bibinfo  {journal} {Phys. Rev. Lett.}\ }\textbf {\bibinfo
  {volume} {75}},\ \bibinfo {pages} {3537} (\bibinfo {year}
  {1995})}\BibitemShut {NoStop}%
\bibitem [{\citenamefont {Schollw\"ock}(2011)}]{dmrg4}%
  \BibitemOpen
  \bibfield  {author} {\bibinfo {author} {\bibfnamefont {U.}~\bibnamefont
  {Schollw\"ock}},\ }\href
  {https://doi.org/http://dx.doi.org/10.1016/j.aop.2010.09.012} {\bibfield
  {journal} {\bibinfo  {journal} {Annals of Physics}\ }\textbf {\bibinfo
  {volume} {326}},\ \bibinfo {pages} {96 } (\bibinfo {year} {2011})},\ \bibinfo
  {note} {january 2011 Special Issue}\BibitemShut {NoStop}%
\bibitem [{\citenamefont {Fubasami}\ \emph {et~al.}(2019)\citenamefont
  {Fubasami}, \citenamefont {Mizoguchi},\ and\ \citenamefont
  {Hatsugai}}]{PhysRevB.100.014438}%
  \BibitemOpen
  \bibfield  {author} {\bibinfo {author} {\bibfnamefont {S.}~\bibnamefont
  {Fubasami}}, \bibinfo {author} {\bibfnamefont {T.}~\bibnamefont
  {Mizoguchi}},\ and\ \bibinfo {author} {\bibfnamefont {Y.}~\bibnamefont
  {Hatsugai}},\ }\href {https://doi.org/10.1103/PhysRevB.100.014438} {\bibfield
   {journal} {\bibinfo  {journal} {Phys. Rev. B}\ }\textbf {\bibinfo {volume}
  {100}},\ \bibinfo {pages} {014438} (\bibinfo {year} {2019})}\BibitemShut
  {NoStop}%
\bibitem [{Note1()}]{Note1}%
  \BibitemOpen
  \bibinfo {note} {In our finite-size calculations the dimerization measured in
  the middle of the chain in the uniflrm phases is $D_\protect \mathrm
  {mid}\propto O(10^{-2})$. For comparison, the mid-chain dimerization in the
  partially dimerized phase in spin-5/2 chain is $D_\protect \mathrm
  {mid}\propto O(1)$}\BibitemShut {NoStop}%
\bibitem [{\citenamefont {Roth}\ and\ \citenamefont
  {Schollw\"ock}(1998)}]{roth}%
  \BibitemOpen
  \bibfield  {author} {\bibinfo {author} {\bibfnamefont {R.}~\bibnamefont
  {Roth}}\ and\ \bibinfo {author} {\bibfnamefont {U.}~\bibnamefont
  {Schollw\"ock}},\ }\href {https://doi.org/10.1103/PhysRevB.58.9264}
  {\bibfield  {journal} {\bibinfo  {journal} {Phys. Rev. B}\ }\textbf {\bibinfo
  {volume} {58}},\ \bibinfo {pages} {9264} (\bibinfo {year}
  {1998})}\BibitemShut {NoStop}%
\bibitem [{\citenamefont {Chepiga}\ \emph
  {et~al.}(2016{\natexlab{b}})\citenamefont {Chepiga}, \citenamefont
  {Affleck},\ and\ \citenamefont {Mila}}]{j1j2j3_long}%
  \BibitemOpen
  \bibfield  {author} {\bibinfo {author} {\bibfnamefont {N.}~\bibnamefont
  {Chepiga}}, \bibinfo {author} {\bibfnamefont {I.}~\bibnamefont {Affleck}},\
  and\ \bibinfo {author} {\bibfnamefont {F.}~\bibnamefont {Mila}},\ }\href
  {https://doi.org/10.1103/PhysRevB.94.205112} {\bibfield  {journal} {\bibinfo
  {journal} {Phys. Rev. B}\ }\textbf {\bibinfo {volume} {94}},\ \bibinfo
  {pages} {205112} (\bibinfo {year} {2016}{\natexlab{b}})}\BibitemShut
  {NoStop}%
\bibitem [{\citenamefont {Di~Francesco}\ \emph {et~al.}(1997)\citenamefont
  {Di~Francesco}, \citenamefont {Mathieu},\ and\ \citenamefont
  {S{\'e}n{\'e}chal}}]{diFrancesco}%
  \BibitemOpen
  \bibfield  {author} {\bibinfo {author} {\bibfnamefont {P.}~\bibnamefont
  {Di~Francesco}}, \bibinfo {author} {\bibfnamefont {P.}~\bibnamefont
  {Mathieu}},\ and\ \bibinfo {author} {\bibfnamefont {D.}~\bibnamefont
  {S{\'e}n{\'e}chal}},\ }\href {https://books.google.ch/books?id=keUrdME5rhIC}
  {\emph {\bibinfo {title} {Conformal Field Theory}}},\ Graduate Texts in
  Contemporary Physics\ (\bibinfo  {publisher} {Springer},\ \bibinfo {address}
  {New York},\ \bibinfo {year} {1997})\BibitemShut {NoStop}%
\bibitem [{\citenamefont {Vanderstraeten}\ \emph {et~al.}(2020)\citenamefont
  {Vanderstraeten}, \citenamefont {Wybo}, \citenamefont {Chepiga},
  \citenamefont {Verstraete},\ and\ \citenamefont
  {Mila}}]{PhysRevB.101.115138}%
  \BibitemOpen
  \bibfield  {author} {\bibinfo {author} {\bibfnamefont {L.}~\bibnamefont
  {Vanderstraeten}}, \bibinfo {author} {\bibfnamefont {E.}~\bibnamefont
  {Wybo}}, \bibinfo {author} {\bibfnamefont {N.}~\bibnamefont {Chepiga}},
  \bibinfo {author} {\bibfnamefont {F.}~\bibnamefont {Verstraete}},\ and\
  \bibinfo {author} {\bibfnamefont {F.}~\bibnamefont {Mila}},\ }\href
  {https://doi.org/10.1103/PhysRevB.101.115138} {\bibfield  {journal} {\bibinfo
   {journal} {Phys. Rev. B}\ }\textbf {\bibinfo {volume} {101}},\ \bibinfo
  {pages} {115138} (\bibinfo {year} {2020})}\BibitemShut {NoStop}%
\end{thebibliography}%

\end{document}